# Oriented artificial nanofibers and laser induced periodic surface structures as substrates for Schwann cells alignment

## Authors and affiliations

Sebastian Lifka[1, *], Cristina Plamadeala[2, *], Agnes Weth[1], Johannes Heitz[2] and Werner Baumgartner[1]

[1] Institute of Biomedical Mechatronics, Johannes Kepler University of Linz, Altenberger Str. 69, 4040 Linz; agnes.weth@jku.at, werner.baumgartner@jku.at

[2] Institute of Applied Physics, Johannes Kepler University of Linz, Altenberger Str. 69, 4040 Linz; johannes.heitz@jku.at

[*] Correspondence: sebastian.lifka@jku.at , cristina.plamadeala@jku.at

## Abstract

People with injuries to the peripheral nervous system, due to its poor functional regeneration, suffer from paralysis of the facial muscles, fingers and hands, or toes and feet, often for the rest of their lives. Therefore, to improve patients' quality of life, there is an urgent need for conduits that effectively support the healing of large defects in nerve pathways through specific guidance of nerve cells. This paper describes two specific methods for achieving directed growth of Schwann cells, a type of glial cells that can support the regeneration of the nerve pathway by guiding the neuronal axons in the direction of their alignment. One method implies the exposure of a poly(ethylene terephthalate) (PET) foil to a KrF* laser beam, that renders a nanorippled surface topography. The other method uses aligned polyamide-6 (PA-6) nanofibers produced via electrospinning on a very fast rotating structured collector, which enables easy nanofiber detachment, without additional effort. Schwann cells growth on these substrates was inspected after one week of cultivation by means of scanning electron microscope (SEM). For both methods we show that Schwann cells grow in a certain direction, predetermined by nanoripples and nanofibers orientation. In contrast, cells cultivated onto unstructured surfaces or randomly oriented nanofibers, show an omnidirectional growth behavior.

## Introduction

One possibility to guide Schwann cells would be the use of natural silk. (Millesi, et al., 2021) and (Stadlmayr, et al., 2023), for example, have recently conducted promising research into the use of natural spider silk in nerve regeneration. Since the harvesting of natural silk can be costly and time-consuming, the use of artificial fibers, such as nanofibers produced using the so-called electrospinning process, would be advantageous. Several processes for the production of aligned nanofibers using electrospinning have already been described in the literature (Han, et al., 2022). Directed nanofibers have also been used as scaffolds in nerve regeneration. (Huang, et al., 2015) used directed cellulose acetate butyrate (CAB) nanofibers for the regeneration of nerves in rats.

A second possibility to guide Schwann cells is to create a substrate with nanoripples, also known as laser-induced periodic surface structures (LIPSS), that can be achieved by short-pulsed laser processing (Bonse, Höhm, Kirner, Rosenfeld, & Krüger, 2017). (Rebollar, et al., 2008) showed that nanoripples on polystyrene (PS) enable directed growth of Chinese hamster ovary (CHO) cells proving that nano topography similar to the nanofeatures found in the extracellular matrix (ECM) provides an important external stimulus and induce an orientation of actin fibers of the cytoskeleton. It has been also demonstrated that laser-induced topography in combination with shear stress affect the adhesion, orientation and elongation of mouse Schwann cells (Babaliari, et al., 2021).

In this work we present two methods which can be used effectively in the future for the production of implants in nerve regeneration. The first method enables the production of aligned nanofibers which can be easily detached from the electrospinning collector as an independent non-woven without additional coating or complex additional work, as it was the case up to now and thus significantly simplifies the manufacturing process. This is achieved by a collector rotating at a very high speed which winds up the electrospun fiber as if on a wire drum. The surface of the collector is structured with a special micro-surface structure, which has already been described in more detail in a previous publication (Lifka, et al., 2023). The surface structure allows the nanofiber non-woven to be removed from the collector easily and without leaving any residue. The second method describes the production of laser-induced periodic surface structures (LIPSS), also known as nanoripples, which, due to their resemblance to the ECM, enable the aligned growth of nerve cells. For both methods we show that, in contrast to randomly oriented fibers or unstructured surfaces, they promote oriented cell growth in a direction

## Methods

### Aligned nanofibers

The most important and novel part of this method is the rotating collector whose surface contains a structure in the µm range. This structure has already been presented in a previous publication and consists in principle of a simple triangular geometry which forms a jagged surface structure. In (Lifka, et al., 2023), this structure was produced on a cylindrical collector with the aid of a thread turning tool in the form of a very fine thread, whereby the direction of preference of the structure is orthogonal to the collector axis. This variant represents the most unfavorable alignment of the structure for this application (the aligned nanofibers must be orthogonal to the direction of the structure in order to guarantee its function). Therefore, the structure must be manufactured in such a way that its orientation is parallel to the cylinder/collector axis. In the sense of simple, fast and cost-effective production, the triangular structure was produced by knurling the cylinder surface with axis-parallel grooves.

For this purpose, a cylindrical collector made of aluminum with a diameter of 40 mm was manufactured and machined on a lathe using a knurling wheel (ZEUS knurling wheel 11 AA 15 x 4 x 4 G7 T=0.3 PM a=90°, Article number: 41013392, Hommel+Keller Präzisionswerkzeuge GmbH, Aldingen, Germany) with the triangular surface structure. The structure therefore has a periodicity and depth of approximately 300 µm with a tip angle of 90°. To enable the collector to rotate, it is supported by two ball bearings (6001-2Z, SKF Österreich AG, Steyr, Austria) to the left and right of the structured surface. The bearing blocks were manufactured using a 3D printer (Photon Mono-X, Hongkong Anycubic Technology Co., Shenzen, China). The resin used for the 3D printed parts was the "Blu" resin from Siraya Tech (CA, USA). The collector is driven by a brushless DC motor (DB28M01, Nanotec Electronic GmbH & Co. KG, Feldkirchen, Germany) via a claw coupling (MJC19-4-A, JD12/19-85B; Ruland Manufacturing Co., Inc., Marlborough, UK). The maximum achievable rotational speed of the collector is approximately 14 meter per second. All CAD data and 2D drawing derivation required for the production of the collector are freely available in a Zenodo repository (Lifka, Plamadeala, Weth, Heitz, & Baumgartner, 2024).

For electrospinning, the rotating drum collector assembly (Figure 1B) was mounted on a custom-made electrospinning setup, as shown in Figure 1A. The setup is based on the classic horizontal electrospinning process in which a polymer solution is conveyed evenly through a thin metal needle with the aid of a syringe pump and applied to a collector in the form of thin nanofibers with the use of a strong electric field. The setup consists of a basic frame made of aluminum profiles which enables precise adjustment of the needle-collector distance. The polymer solution is conveyed evenly through a thin metal needle (Sterican 21G x 7/8″ blunt, B. Braun SE, Melsungen, Germany) using a custom-made syringe pump. The necessary electric field is generated by a high-voltage generator (HCP 35-35,000, FuG Elektronik GmbH, Schechen, Germany). The needle is the positive electrode and the collector is the negative electrode. The needle is contacted directly via a crocodile clip, while the collector is contacted via the conductive ball bearings to avoid sliding contact. The polymer solution for the nanofibers consists of 15 wt.% polyamide-6 (PA6) (provided by Elmarco s.r.o., Liberec, Czech Republic) dissolved in formic acid (FA) (ROTIPURAN® ≥98%, p.a., Carl Roth GmbH + Co. KG., Karlsruhe, Germany) and acetic acid (AA) (ROTIPURAN® 100%, p.a., Carl Roth GmbH + Co. KG., Karlsruhe, Germany) in a (weight) ratio of 2:1. The parameters for the spinning process were selected as follows: Needle-collector distance of 13 cm, needle-collector voltage of 20 kV, flow rate of 0.34 mLh$^{-1}$.

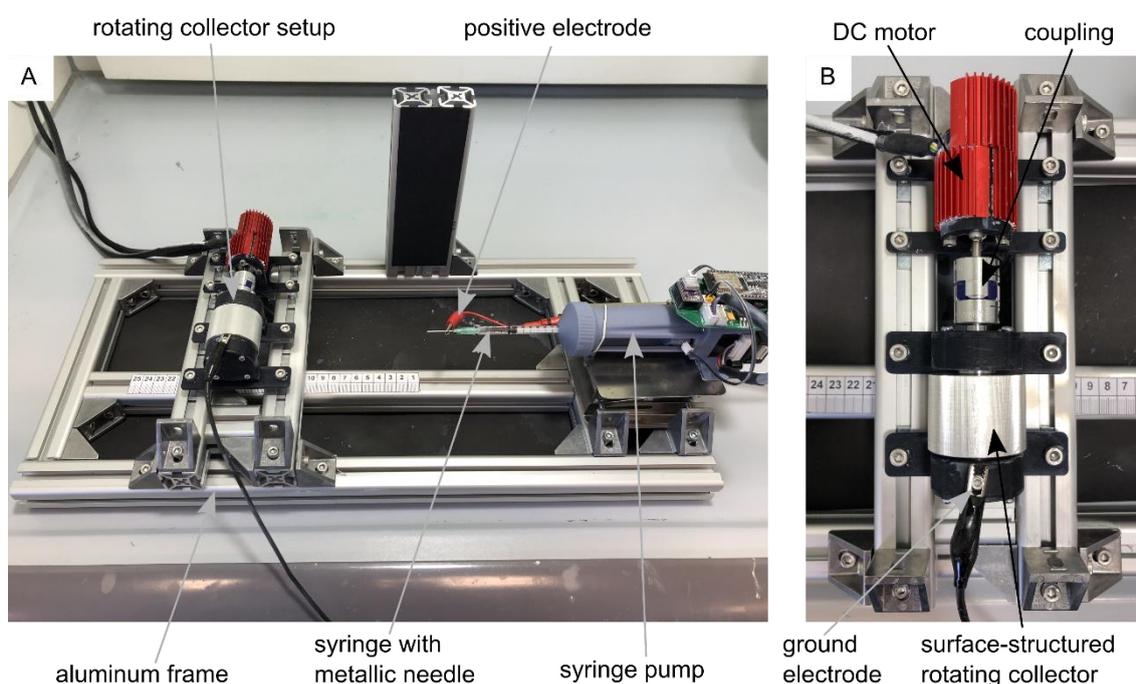

**Figure 1. Electrospinning-setup for the production of aligned nanofibers.** (A) Custom-made electrospinning setup. (B) Rotating drum collector setup mounted on the electrospinning setup-

The motor speed was controlled and set via the software "Plug & Drive Studio 3" (Nanotec Electronic GmbH & Co. KG, Feldkirchen, Germany). Various rotational speeds of the collector were tested until the fibers were oriented in the desired direction. The electrospinning process took about 10 minutes for each sample. After this time, a fleece that was thick and robust enough for further processing could be guaranteed. The finished non-woven was then cut open with a scalpel parallel to the collector axis and carefully removed from the collector with tweezers. Thanks to the surface structuring, the fleece could be removed easily and without leaving any residue. The Zenodo repository (Lifka, Plamadeala, Weth, Heitz, & Baumgartner, 2024) contains a video (Non-Woven_Detachement_from_Collector.mkv) showing the removal process of the aligned nanofiber non-woven. The specimens of the nanofiber non-woven were then sputter-coated with gold (SCD 005, BAL-TEC Inc., Balzers, Liechtenstein) for 80 s with 21 mA and examined under a scanning electron microscope (SEM) (Philips 525, Philips Electron Optics, Amsterdam, The Netherlands). The SEM images subsequently serve as the basis for the analysis of the directionality and the average fiber diameter.

The statistical analyses of fiber directionality and fiber diameter were performed using ImageJ (v 1.51w). The directionality analysis function (v2.3.0, method: Fourier components) was used to analyze the fiber orientation. The DiameterJ 1-018 plug-in was used to analyze the fiber diameter. The corresponding diagrams were created with the help of Python 3 and matplolib from the .csv files generated by ImageJ. All underlying data can be found in the Zenodo repository (Lifka, Plamadeala, Weth, Heitz, & Baumgartner, 2024).

### Nanoripples

To produce ripples on poly(ethylene terephthalate) (PET) foils with a thickness of 50 μm (Goodfellow Ltd., Bad Nauheim, Germany), a KrF* (krypton fluoride) excimer laser (LPX 300, Lambda Physik, 181 Göttingen, Germany) was used, with a wavelength of 248 nm, 20 ns pulse duration, and 10 Hz pulse repetition rate. The setup is shown in Figure 2 and described in more details in (Richter et al., 2021). Prior to exposure, the PET foils were rinsed with ethanol and thoroughly dried with nitrogen, in order to remove any debris. Later on, the foils were fixed onto the sample holder at a 30° incidence angle and exposed to 6000 pulses at an energy of 12.6 – 12.8 mJ, applying an average fluence of 10 mJ cm$^{-2}$. Laser beam energy was measured with a high-area high-damage energy sensor (J45LP-MUV, Coherent, Portland, United States) with the help of a laser energy meter (FieldMaxII-P™, Coherent, Portland, United States).

Scanning electron microscope (model REM 1540XB-Crossbeam, Zeiss, Oberkochen, Germany) images of nanoripples were acquired at ten different positions within the nanostructured area, and analyzed by using the free software Gwyddion (version 2.61, Czech _Metrology Institute, Brno, Czech Republic) in order to calculate the spatial period. The underlying data for spatial period calculation can be found in the Zenodo repository (Lifka, Plamadeala, Weth, Heitz, & Baumgartner, 2024).

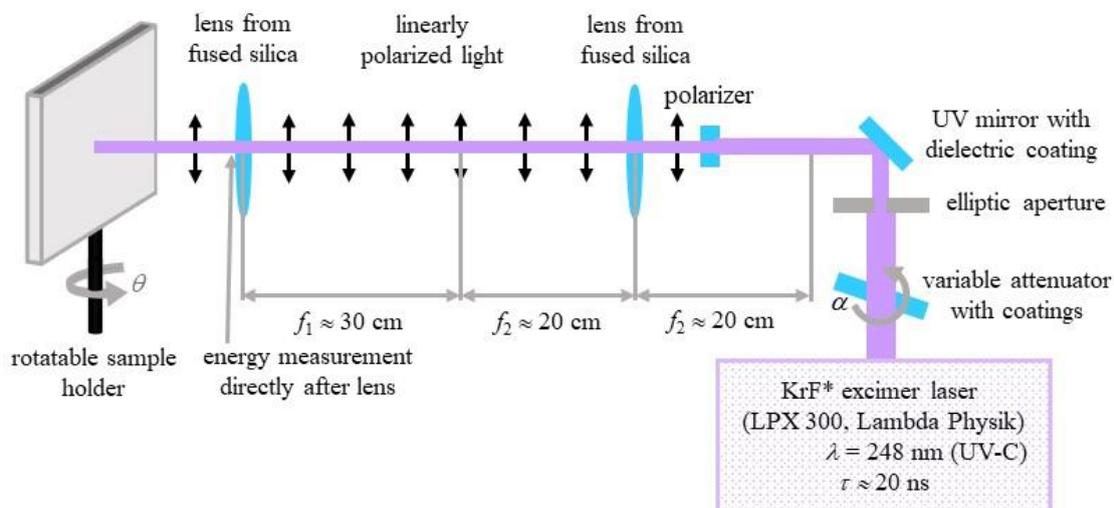

**Figure 2. Setup for laser-induced periodic surface structures (LIPSS) fabrication on poly(ethylene terephthalate) (PET) foils.** (Reprinted from (Buchberger, et al., 2023) which is an open-access article distributed under the terms of the Creative Commons Attribution 4.0 International (CC-BY 4.0))

### Cell cultivation

To validate and show the oriented cell growth on samples produced with the above described methods murine Schwann cells (Immortalized Mouse Schwann Cells (IMS32), T0295, Applied Biological Materials Inc., Richmond, Canada) were cultured for one week in PriGrow III (TM003, Applied Biological Materials Inc., Richmond, Canada) + 10% fetal bovine serum (FBS) + 1% Penicillin/Streptomycin Solution (G255, Applied Biological Materials Inc., Richmond, Canada) medium at 37.0 °C and 5% $CO_2$ in a New Brunswick Galaxy® 48 S CO2 incubator (Eppendorf, Hamburg, Germany) on the corresponding sample.

## Results

### Aligned nanofibers

Figure 3A to C show an example of the detachment process for the nanofiber non-woven. It can be seen that the non-woven can be removed easily and without leaving any residue on the collector. The video "Non-Woven_Detachement_from_Collector.mkv" in the Zenodo repository (Lifka, Plamadeala, Weth, Heitz, & Baumgartner, 2024) shows the removal process again in detail.

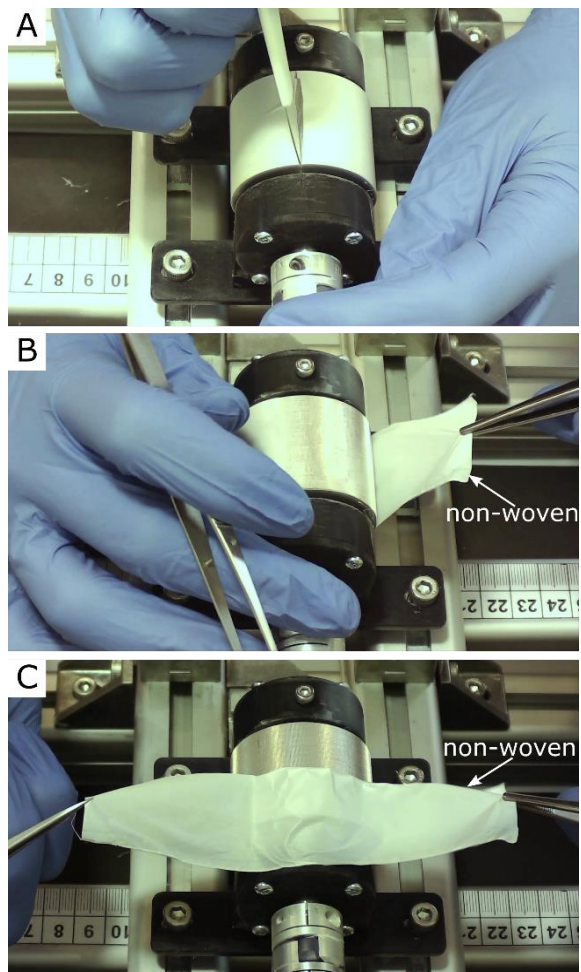

**Figure 3. Removal process of the oriented nanofiber non-woven.** (A) Cut open the fleece with a scalpel. (B) Remove the fleece with tweezers. (C) Fleece completely detached from the collector without leaving any residue on the collector.

Figure 4 shows the SEM images of four different nanofiber non-woven. Figure 4A shows a non-woven with randomly oriented fibers (i.e. the collector did not rotate) as a control. Figure 4B to C show non-woven with increasing collector rotation speed. It can be clearly seen that the degree of orientation of the fibers increases with the rotational speed of the collector. While the orientation of the fibers still has potential for improvement at a rotational speed of $v = 5$ ms$^{-1}$ (Figure 4B), a clear orientation in the horizontal direction can already be seen at a rotational speed of $v = 8$ ms$^{-1}$ or $v = 14$ ms$^{-1}$ (Figure 4C and D).

To better quantify the directionality of the fibers, Figure 5 shows the directionality histograms for all SEM images of Figure 4. The horizontal line in the histogram represents an angle of 0°. The dashed curve in red represents a normal distribution fit. It can be clearly seen that the sample with the randomly oriented fibers (Figure 5A) is relatively evenly distributed in all directions. The sample with $v = 5$ ms$^{-1}$ (Figure 5B) is much more directional in comparison. The samples in which the collector was rotated at $v = 8$ ms$^{-1}$ (Figure 5C) and $v = 14$ ms$^{-1}$ (Figure 5D) both exhibit excellent directionality.

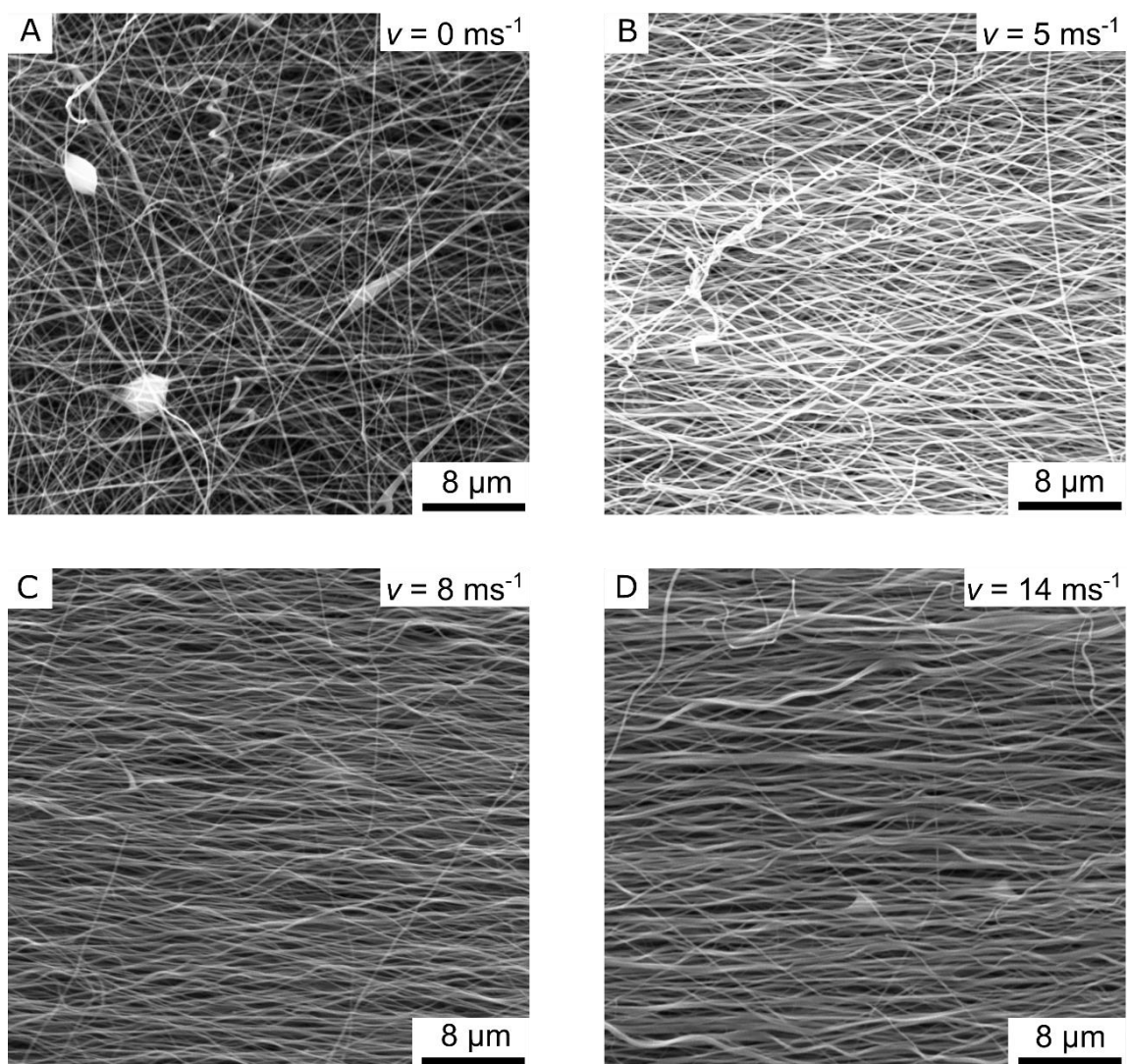

**Figure 4. SEM images of the aligned nanofiber non-woven.** (A) Randomly oriented nanofibers as control (i.e. the collector did not rotate, $v = 0$ ms$^{-1}$). (B) Nanofibers collected on the collector rotating with $v = 5$ ms$^{-1}$. (C) Nanofibers collected on the collector rotating with $v = 8$ ms$^{-1}$. (D) Nanofibers collected on the collector rotating with $v = 14$ ms$^{-1}$.

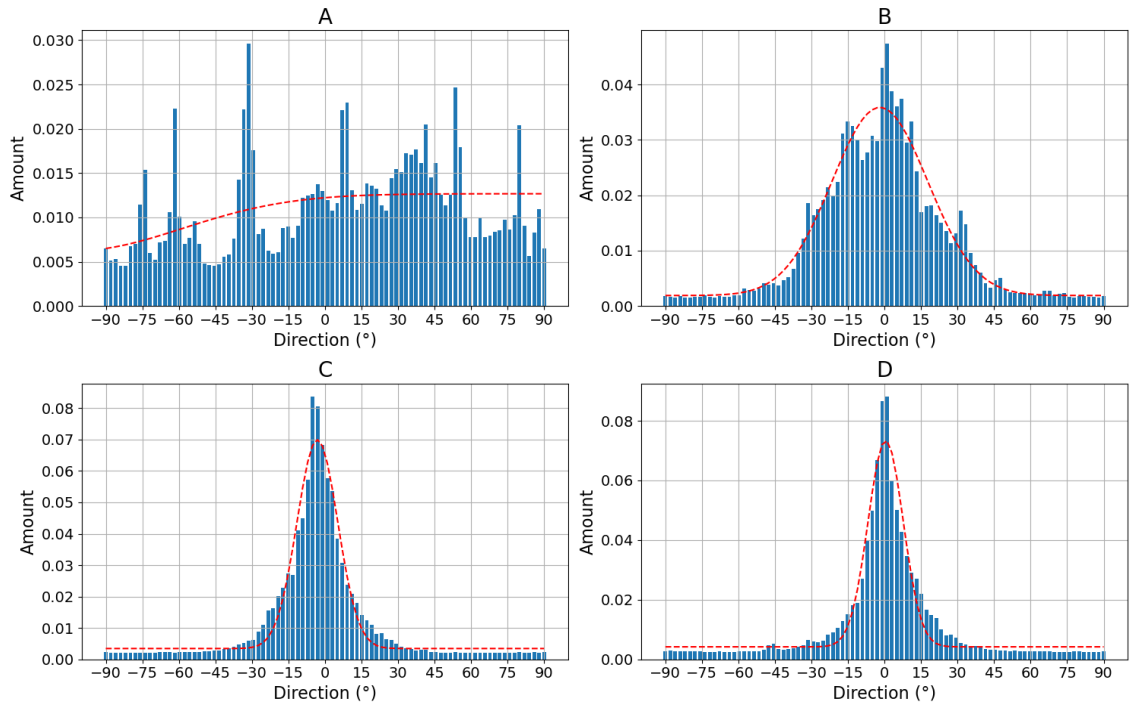

**Figure 5. Directionality histograms of the non-woven.** (A) Randomly oriented nanofibers as control (i.e. the collector did not rotate, $v = 0$ ms$^{-1}$). (B) Nanofibers collected on the collector rotating with $v = 5$ ms$^{-1}$. (C) Nanofibers collected on the collector rotating with $v = 8$ ms$^{-1}$. (D) Nanofibers collected on the collector rotating with $v = 14$ ms$^{-1}$. The dashed red line represents a normal distribution fit of the histogram.

Table 1 lists the average fiber diameters of the individual samples. On average across all samples, the fibers produced had a diameter of approx. 250 nm with a standard deviation of approximately 110 nm.

**Table 1. Mean fiber diameters of the samples.**

|  | Mean fiber diameter in nm | Standard deviation in nm |
|---|---|---|
| **Random oriented** | 277.5 | 110.4 |
| **5 ms$^{-1}$** | 205.9 | 81.3 |
| **8 ms$^{-1}$** | 226.5 | 101.4 |
| **14 ms$^{-1}$** | 309.0 | 151.5 |
| **Overall** | 254.725 | 111.15 |

To investigate the directional cell growth on the samples, murine Schwann cells were cultured on an 8 ms$^{-1}$ non-woven nanofiber sample (Figure 4C), which has a parallel fiber direction (Figure 5C). The result can be seen in Figure 6A. A clear orientation of the cell growth in a horizontal direction can be seen. Figure 6B shows an enlarged section of Figure 6A, in which the non-woven under the cells is also visible. Note that the orientation of the nanofibers is also horizontal (i.e. parallel to the direction of cell growth).

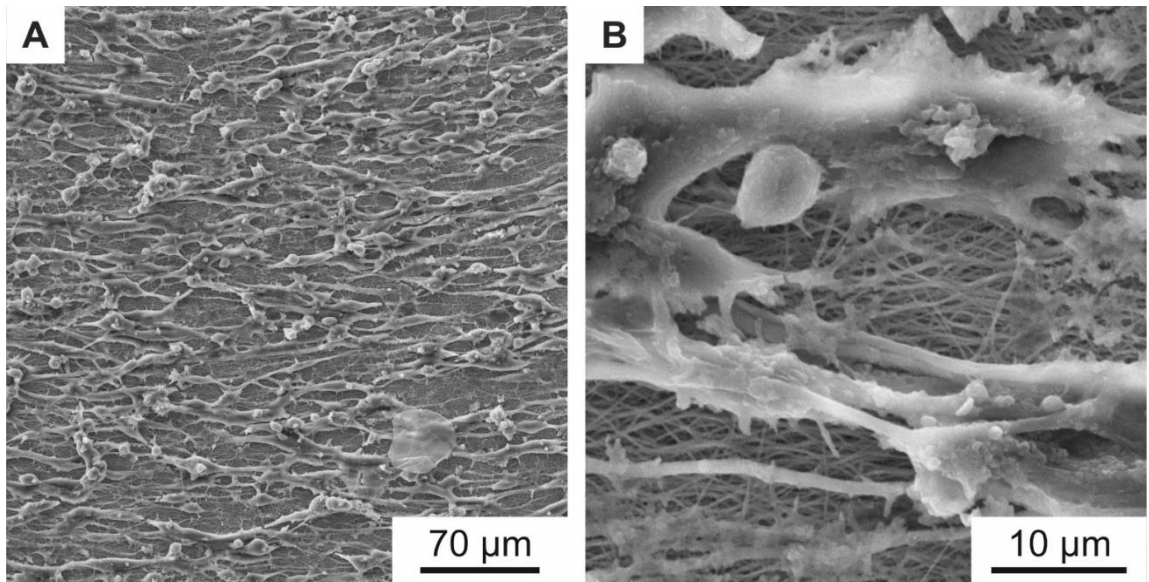

**Figure 6. SEM images of aligned murine Schwann cells cultivated for one week on electrospun PA-6 nanofibers with preferential orientation.** (B) is a magnified detail of (A), where the underlying nearly parallel oriented nanofibers are better visible. The nanofiber sample was spun with a rotational collector speed of 8 ms$^{-1}$ as shown in Figure 4C.

### Nanoripples

Nanoripples (LIPSS) are structures that form upon laser irradiation on many materials, and originate from the interference of the incoming linearly polarized light of (ultra-)short pulsed lasers with the radiation remnants or plasmon modes in the surface (Bonse, Höhm, Kirner, Rosenfeld, & Krüger, 2017). The orientation of the LIPSS can be either parallel or perpendicular to the polarization of the laser beam, and their periodicity is comparable to the wavelength or smaller, depending on the laser fluence and the angle of incidence of the laser beam onto the irradiated substrate. In our case, the spatial period $\Lambda$ of nanoripples fabricated using s-polarized laser light is given by $\Lambda = \lambda/(n_{eff} - \sin\theta)$, where $\lambda$ is the wavelength of the laser beam, $n_{eff}$ – the effective refractive index which lies between the refractive indices of air and PET (Barb, et al., 2014), and $\theta$ – the angle of incidence.

Figure 7A shows the murine Schwann cells cultivated for one week (same procedure as described above) on flat PET foil, used as reference sample. The cells show a random orientation and omnidirectional growth. In contrast, Figure 7B, showing murine Schwann cells cultivated for one week on PET nanoripples, proves the influence of nanorippled topography as external stimuli that renders cell alignment. Schwann cells exhibit a typical elongated bipolar shape and even though their sizes (in the order of some tens of micrometers) are much bigger than the nanoripples' height (above 110 nm (Buchberger, et al., 2023)) and periodicity ($\Lambda = 331 \pm 11$ nm, $N = 10$), the cells align and orient themselves along the LIPSS orientation. This effect happens due to the interaction of cell filopodia with the nano topography, as shown in Figure 7C and D.

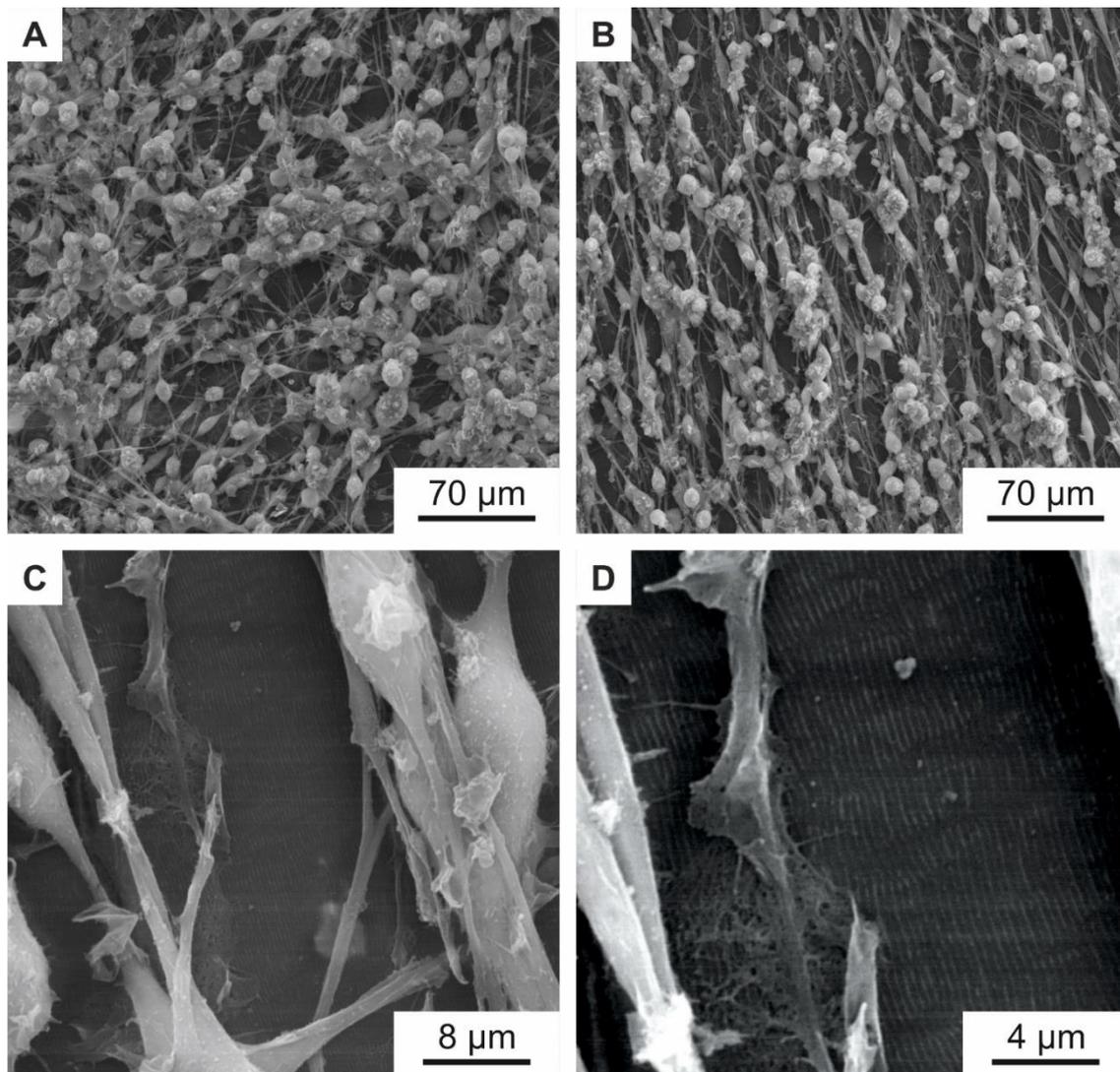

**Figure 7. Murine Schwann cells on PET foils.** (A) SEM image of murine Schwann cells cultivated for one week on flat PET foil, as control, showing no alignment. (B-D) SEM images of aligned murine Schwann cells cultivated for one week on a PET foil with nanoripples. (C, D) are magnifications of (B), showing an area with incomplete cell coverage, proving that the cells and the nanoripples have the same orientation.

## Discussion

In this work, two methods are described to present an improved directed growth of Schwann cells, a type of glial cell that insulates and protects the axons of neurons. Major injuries of nerve tissue are often characterized by poor functional regeneration and therefore require conduits that promote healing of the nerve axons and provide direction for Schwann cell growth. One method uses laser-induced periodic surface structures (LIPSS) that mimic the nanofeatures found in the extracellular matrix. These LIPSS can be used to specify the direction of growth for Schwann cells and promote growth in a specific direction. However, this is limited to more or less rigid (in this case) PET films. Therefore, a second method describes the production of a non-woven consisting of nanofibers. This non-woven is extremely flexible and can be adapted to many different shapes. The method is based on the electrospinning process and uses a very fast rotating cylindrical collector to produce directed nanofibers. The special feature of the method is the special surface structure (Lifka, et al., 2023) of the collector, which enables the non-woven to be easily removed without additional coating and additional effort. In this work, only one fiber material was tested, namely PA-6, as good experience has already been made with this polymer. Of course, this method can also be applied to biocompatible polymers, such as fibers made of cellulose acetate butyrate (CAB) or collagen fibers. These materials would be very well suited for use in nerve regeneration.

## Conclusion

In this paper we show that oriented mechanical nanofeatures, such as nanofibers and nanoripples, are able to mimic the extracellular matrix features, and stimulate the elongation and support the alignment of Schwann cells. These attributes may subsequently lead to a better axonal regeneration.

## Data and software availability
**Underlying data**

Zenodo: Repository for the Method Article "Oriented artificial nanofibers and laser induced periodic surface structures as substrates for Schwann cells alignment". https://doi.org/10.5281/zenodo.10665997.

This project contains the following underlying data:

- Directionality_Hist.py. (Python file for generating directionality histograms of the non-woven.)
- RandomHist.csv. (Histogram data from ImageJ for random oriented non-woven.)
- 5mpsHist.csv. (Histogram data from ImageJ for 5 ms$^{-1}$ non-woven.)
- 8mpsHist.csv. (Histogram data from ImageJ for 8 ms$^{-1}$ non-woven.)
- 14mpsHist.csv. (Histogram data from ImageJ for 14 ms$^{-1}$ non-woven.)
- DiameterAnalysisRandom.csv. (Fiber diameter analysis and histogram data from ImageJ for random oriented non-woven.)
- DiameterAnalysis5mps.csv. (Fiber diameter analysis and histogram data from ImageJ for 5 ms$^{-1}$ non-woven.)
- DiameterAnalysis8mps.csv. (Fiber diameter analysis and histogram data from ImageJ for 8 ms$^{-1}$ non-woven.)
- DiameterAnalysis14mps.csv. (Fiber diameter analysis and histogram data from ImageJ for 14 ms$^{-1}$ non-woven.)
- Spatial period calculation.docx. (Word file containing the data for the spatial period calculation of the nanoripples.)

**Extended data**

Zenodo: Repository for the Method Article "Oriented artificial nanofibers and laser induced periodic surface structures as substrates for Schwann cells alignment". https://doi.org/10.5281/zenodo.10665997.

This project contains the following extended data:

- Bearing bracket Fixed bearing.step. (CAD .step file for the fixed bearing bracket of the rotating drum collector setup.)
- Bearing bracket loose fit bearing.step. (CAD .step file for the loose bearing bracket of the rotating drum collector setup.)
- Bearing cover Fixed bearing.step. (CAD .step file for the fixed bearing cover of the rotating drum collector setup.)
- Bearing cover loose fit bearing.step. (CAD .step file for the loose bearing cover of the rotating drum collector setup.)
- Motor mount.step. (CAD .step file for the DC motor mount of the rotating drum collector setup.)
- Surface structured collector.step. (CAD .step file for the surface-structured rotating drum collector.)
- Surface Structured Collector Drawing.pdf. (2D drawing of the rotating drum collector.)
- Setup Rotating Collector.pdf. (2D drawing of the whole rotating drum collector setup.)
- Non-Woven_Detachement_from_Collector.mkv (Short video showing the detachment process of the non-woven fabric from the surface-structured collector.)

## Competing interests
No competing interests were disclosed.

## Grant information
This work was financially supported by the European Union's Horizon 2020 research and innovation program under the grant agreement No [862016] (project [BioCombs4Nanofibers], Johannes Heitz and Werner Baumgartner) and by the Austrian COMET-K2 program of the Linz Center of Mechatronics (LCM).

## Acknowledgements
We would like to thank ZONA for their support and assistance in the production of the surface-structured collector for the electrospinning process.